\documentclass[a4paper,10pt,pre,showpacs,aps,floats,superscriptaddress]{revtex4}

\usepackage{epsfig}\usepackage{amsmath}

\newcommand{\avk}{\langle k \rangle}

\begin{document}

\title{Fitness for Synchronization of Network Motifs} 

\author{Yamir Moreno}

\affiliation{Departamento de F\'{\i}sica Te\'orica, Universidad de
Zaragoza, Zaragoza 50009, Spain}

\affiliation{Instituto de Biocomputaci\'on y F\'{\i}sica de Sistemas
Complejos, Universidad de Zaragoza, Zaragoza 50009, Spain}

\author{Miguel V\'azquez-Prada}

\affiliation{Departamento de F\'{\i}sica Te\'orica, Universidad de
Zaragoza, Zaragoza 50009, Spain}

\author{Amalio F. Pacheco} 

\affiliation{Departamento de F\'{\i}sica Te\'orica, Universidad de
Zaragoza, Zaragoza 50009, Spain}

\affiliation{Instituto de Biocomputaci\'on y F\'{\i}sica de Sistemas
Complejos, Universidad de Zaragoza, Zaragoza 50009, Spain}

\date{\today}

\widetext

\begin{abstract} 

We study the phase synchronization of Kuramoto's oscillators in small
parts of networks known as motifs. We first report on the system
dynamics for the case of a scale-free network and show the existence
of a non-trivial critical point. We compute the probability that
network motifs synchronize, and find that the fitness for
synchronization correlates well with motif's interconnectedness and
structural complexity. Possible implications for present debates about
network evolution in biological and other systems are discussed.

\end{abstract}


\maketitle

\section{Introduction}

During the last years, many scientists have scrutinized the world
around us looking for regularities. One of the most recent findings is
the fact that many real systems can be cast into a common topological
structure, called complex networks \cite{strogatz}. Examples include
biological \cite{ref5,ref6} social \cite{pnas} and technological
\cite{alexei} systems that exhibit similar patterns of
interconnections \cite{strogatz}. They are characterized by the
existence of key elements in the network which drastically reduce the
average distance between all of them, the so-called small-world
property \cite{ws98}. Additionally, it turns out that for a large
number of real-world systems, the probability that any given element
(node) of the system interacts with (is linked to) $k$ other
components, follows a power-law $P(k)\sim k^{-\gamma}$, with and
exponent $\gamma$ usually estimated between $2$ and $3$. These
networks have been termed scale-free (SF) networks.

The above property was soon shown to be at the root of distinct
behaviours when several dynamical processes are placed on top of SF
networks. These are the cases of percolation and epidemic spreading
processes, intensively studied during the last years due to their
practical relevance in different applications and the availability of
analytical treatments
\cite{newman00,pv01a,moreno02,av03,moreno03}. The peculiar topological
properties of the underlying network for these two processes lead to
the absence of any percolation or spreading threshold in the
thermodynamic limit, a previously well-established result for regular
and random graphs.

It is then natural to ask whether or not and to what extent the
topology of complex networks influences the behaviour of other dynamical
processes. In particular, for biological and other applications, it
would be relevant to consider the nodes of a given network as
nonlinear dynamical systems. The behaviour of an isolated generic
dynamical system in the long-term limit can be described by stable
fixed points, limit cycles or chaotic attractors
\cite{sbook}. However, we have learned in recent years that when many
of such dynamical systems are coupled together, the details matter. In
this way, the study of networks with both dynamical and structural
complexity might shed light on a number of relevant open problems
where nonlinearity and spatial complexity coexist. Dynamical
complexity may manifest itself through self-organization,
synchronization, the emergence of order, etc.

In this paper, we study the emergence of collective phase
synchronization \cite{pik} in scale-free networks and in simple
topological configurations such as triangles, squares and pentagons
with a variable number of internal connections, called network
motifs. To this end, we study the model proposed by Kuramoto several
years ago and show how a large number of the system's constituents
forms a common dynamical pattern, despite the intrinsic differences in
their individual dynamics. On a theoretical level, we point out that
the fitness for synchronization of network motifs correlates well with
their conservation in the evolution of protein motifs \cite{ref3}, which
might hint at new connections between graph theory, dynamics on
networks and biological systems.

\section{The Kuramoto Model in Scale-Free Networks}

In order to study how topology influences collective dynamics, we
assume that each network's component is an oscillator and that each
interacts with the others following the Kuramoto model
\cite{k1,k2}. This choice is due to the remarkable prestige achieved
by synchronization as a cooperative phenomenon within nonlinear
science and to the seniority, elegance and universality of this
model. Specifically, each node $i$ is considered to be a planar rotor
characterized by an angular phase, $\theta_i$ , and a natural
frequency $\omega_i$. Linked nodes interact with a coupling strength
$\lambda$ according to
\begin{equation}
\frac{d\theta_i}{dt}=\omega_i+\lambda\sum_j^{k_i}\sin(\theta_j-\theta_i)
\label{eq1}
\end{equation}
where $k_i$ is the number of neighbors of the rotor $i$ as given by
the actual architecture of the underlying graph. For the present work,
the natural frequencies and the initial values of $\theta_i$ have been
randomly drawn from a uniform distribution $g(\omega)$ with mean
$\omega_0=0$ in the interval $(-1/2,1/2)$ and $(-\pi,\pi)$,
respectively. Synchronization occurs when $\lambda$ exceeds a critical
value, at which clusters of frequency-locked oscillators appear. This
state represents the emergence of cooperation between network's
constituents.

\begin{figure}[t]
\begin{center}
\epsfig{file=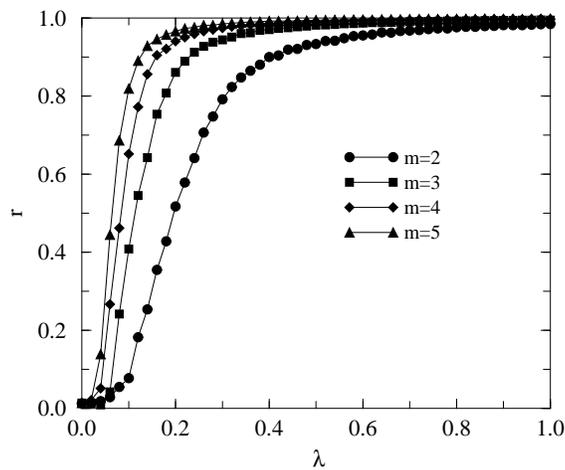,width=2.5in,angle=-90,clip=1}
\end{center}
\caption{Phase diagram of the Kuramoto model in random scale-free
  networks for several average connectivity values $\avk=2m$. The
  onset of synchronization occurs at a nonzero value of $\lambda$ in
  all cases. Although not shown, the transition follows the mean-field
  behaviour exhibited by the globally coupled map. Each value of $r$ is
  the result of at least 10 network realizations and 1000 iterations
  for $N=10^4$ nodes.}
\label{figure1}
\end{figure}

The original Kuramoto model corresponds to the simplest case of
globally coupled (all-to-all), equally weighted oscillators where the
coupling strength $\lambda=K/N$ to ensure that the model is well
behaved in the thermodynamic limit \cite{k1,k2}. For this model and
without any interaction, $K=0$, the oscillators follow their own
dynamics as determined by their natural frequencies and thus the
system is unable to synchronize. However, as $K$ increases the
population of rotors becomes more coherent and few oscillators form a
small cluster of ordered (synchronized) states. If $K$ is further
increased, the synchronized pack tends to recruit more and more
oscillators and eventually the system settles into a unique and
totally synchronous state. The onset of synchronization occurs at a
critical value of the coupling strength, $K_c=2/\pi g(\omega_0)$. The
second-order phase transition is characterized by the order parameter
\begin{equation}
r(t)=\left|\frac{1}{N}\sum_{j=1}^{N}e^{i\theta_j(t)}\right|
\label{eq2}
\end{equation}
which behaves when both $N\rightarrow\infty$ and $t\rightarrow\infty$
as $r\sim (K-K_c)^{\beta}$ for $K\ge K_c$ being $\beta=1/2$.

In order to study the dynamics of the model on top of complex
heterogeneous networks, we first generated SF nets using the BA
procedure \cite{bar99}. In this model, starting from a set of $m_0$
nodes, one preferentially attaches each time step a newly introduced
node to $m$ older ones. The procedure is repeated $N-m_0$ times
and a network made up of $N$ nodes with a power law degree
distribution $P(k)\sim k^{-\gamma}$ with $\gamma=3$ and average
connectivity $\langle k \rangle=2m$ builds up. This network is a clear
example of a highly heterogenous network in that the degree
distribution has unbounded fluctuations when $N\rightarrow\infty$.

We have performed extensive simulations of the model \cite{us} through
numerical integration of the equations of motion Eq.\ (\ref{eq1}). In
the case of random SF networks the global dynamics of the system is
qualitatively the same as for the original Kuramoto model. The phase
diagram of the system is shown in Fig.\ \ref{figure1} for a network of
$N=10^4$ nodes and several values of the average connectivity
$\avk=2m$, where $r$ is the order parameter as given by Eq.\
(\ref{eq2}). Starting from small values of the coupling strength, the
interactions do not overcome the tendency of each rotor to oscillate
according to its individual dynamics. In this state, the behaviour of
the system is completely incoherent and no synchronization is
achieved. This picture persists until a certain critical value
$\lambda_c$ is crossed. At this point some elements lock their phase
and a cluster of synchronized nodes comes up. This constitutes the
onset of synchronization. Beyond this value, there are several groups
or clusters within which the nodes are either synchronized and locked
in phase or still governed by their intrinsic dynamics and thus in an
asynchronous state. The first groups add to $r$ and it departs from
zero as $\lambda$ is increased beyond $\lambda_c$. Finally, after
further increasing the value of $\lambda$, more and more nodes cluster
around the mean phase and the system eventually settles into a
completely synchronized state where $r\approx1$.

On the other hand, Fig.\ \ref{figure1} also provides evidence that the
critical point at which synchronization is attained in these SF
networks is not zero \cite{us}. This is more clearly appreciated if
one eliminates the dependency observed in the figure of the parameter
$r$ on the average connectivity of the network $\avk$. If in Eq.\
(\ref{eq1}) the term $\lambda$ is substituted by $\lambda/\avk$, the
$\avk$-dependency is avoided and all curves in Fig.\ \ref{figure1}
collapse into a single one, showing a critical coupling strength
$\lambda_c=0.25(3)$ signaling that even for large values of the
average connectivity, $\lambda_c$ is not equal to zero. Additionally,
finite-size scaling analysis \cite{us} shows that the transition
remains of the second-order type as for the case of fully connected
networks.

\section{Synchronization of Network Motifs}

Let's now turn our attention to small geometrical structures known as
motifs \cite{milo1,milo2,cap} instead of looking at the whole
network. These structures can be defined as graph components which are
observed in a given network more frequently than in a completely
random graph with identical $P(k)$. Triangles and rectangular loops
are among these graph components, also known as cycles. They are
important because they express the degree of redundancy and
multiplicity of paths among nodes in the topology of the network and
reveal the existence of hierarchical levels. Hence, the extension of
the study of synchronization phenomena to motifs is also relevant.

\begin{figure}[t]
\begin{center}
\epsfig{file=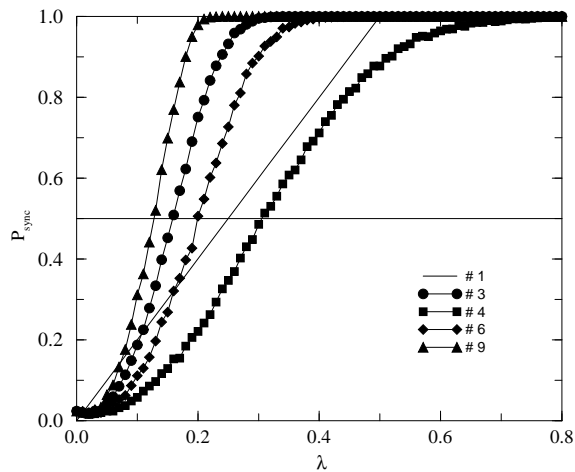,width=2.5in,angle=-90,clip=1}
\end{center}
\caption{Probability of synchronization for five topological
  motifs. Configurations are according to Table\ \ref{table1}. Each
  value of $P_{sync}$ is the result of 1000 iterations over the
  distribution of $\omega_i$ for each value of $\lambda$. The
  horizontal line marks the threshold of $P_{sync}$ for the
  computation of $\lambda^*$. See the text for further details.}
\label{figure2}
\end{figure}

We have computed for all motifs up to $N=4$ nodes the probability,
$P_{sync}$, that synchronization occurs. This probability is an
increasing function of $\lambda$, and is calculated for each $\lambda$
by randomly drawing the values of the natural frequencies $\omega_i$
from a uniform distribution in the interval $(-1/2,1/2)$. Starting
from $\lambda=0$, one averages, for every $\lambda$, over many
realizations of $\omega_i$ and computes the relative number of
simulations where synchronization is accomplished, $P_{sync}$. We
define $\lambda^*$, which varies from motif to motif, as the value of
$\lambda$ beyond which $P_{sync}\ge \frac{1}{2}$. The lower
$\lambda^*$ is for a motif, the better it synchronizes. In Fig.\
\ref{figure2}, we have represented $P_{sync}$ as a function of
$\lambda$ for several motifs.

The simplest case of the dimer ($\#1$ in the Table) can be deduced
analytically. Using the same notation as in Eq. (\ref{eq1}), here we
have:

\begin{equation}\label{eq3}
\frac{d \theta_1}{d t}= \omega_1 + \lambda \ \sin{( \theta_2
-\theta_1)}, \
  \frac{d \theta_2}{d t}= \omega_2 + \lambda \ \sin{(
\theta_1 -\theta_2)}.
\end{equation}

To solve these equations, we introduce the new variables,

\begin{equation}\label{eq5}
f=\theta_2 -\theta_1 , \ \kappa=\omega_2 -\omega_1,\ c=\frac{2
\lambda}{\kappa}.
\end{equation}

Thus, $f$ is the phase difference between the rotors, $\kappa$ is the
difference in the natural frequencies and $c$ is a dimensionless
constant equal to twice the ratio $\lambda/\kappa$. From Eq.\
(\ref{eq3}) we obtain,

\begin{equation}\label{eq7}
\frac{d f}{\kappa(1- c\; \sin{f})}= dt\ .
\end{equation}

The integration of Eq. (\ref{eq7}) leads to three solutions depending
on the value of $c$. The three distinct results of the integral in
Eq. (\ref{eq7}) indicate a different functional behaviour of the
system when $c<1$, $c=1$, and when $c>1$. In fact, $c=1$ is the
critical point. In consequence, from the definition of $c$ in
Eq. (\ref{eq5}), the critical coupling of this system corresponds to

\begin{equation}\label{eq11}
 \lambda^*= \frac{\kappa}{2}=(\omega_2-\omega_1)/2 \ .
\end{equation}

Recalling that the $\omega$s are drawn from a uniform probability
distribution in the interval $(-1/2,1/2)$ with a mean equal to zero,
i.e., $\omega_1=-\omega_2=\omega$, we have that the synchronization
condition reduces to $\lambda\ge\omega$. The probability $P_{sync}$ is
then equal to $P_{sync}=\frac{\lambda}{1/2}=2\lambda$, for $\lambda <
\frac{1}{2}$, and $P_{sync}=1$ for $\lambda \ge \frac{1}{2}$. Thus,
for the dimer we have $\lambda^*=\frac{1}{4}$. For the other motifs,
$P_{sync}$ has been calculated numerically by means of a
$4^{th}$-order Runge-Kutta method. The results are summarized in Fig.\
\ref{figure2} for several motifs (see also Table\ \ref{table1} below).

From Fig.\ \ref{figure2}, one realizes that for a fixed motif's size,
the way in which connections are established determines the
synchronization threshold of that motif configuration. For instance,
configurations $4$, $6$, and $9$ correspond to motifs made up of four
oscillators with a variable number of connections among them. However,
their $\lambda^*$ are quite different. This indicates that the details
of the local and internal connections matter. In particular, we
observe that the higher the interconnection between motif's
constituents, the lower $\lambda^*$.

\section{Discussions and Conclusions}

We now show how the present study may provide useful hints to a
better understanding of several experimental observations in biological
systems. Recent advances in network analysis and modeling have
provided a promising approach to understanding basic biological
organization \cite{ref1,ref2}. In this context, quantitative evidence
has been recently reported \cite{ref3} that in {\em Saccharomyces
cerevisiae}, the proteins organized in definite cohesive patterns of
interaction, and these patterns themselves, are conserved in the
evolution across species in a substantially higher degree than those
that do not participate in such specific motifs. A second finding
\cite{ref3} is that the conservation of proteins in distinct
topological motifs correlates with their interconnectedness.

These observations take place in the context of activities occurring
within a living cell and which no doubt, are of extreme
complexity. Non-linear dynamics prevails at this level of
organization, where units that interact according to simple rules can
generate unexpected complex patterns \cite{sbook}. On a theoretical
level, one may hypothesize that the observed persistence across
evolution is due to some mechanism aimed at optimizing the cooperation
between neighboring nodes. A first approach would then be to see if
the same structures favor the synchronization as a cooperative phenomenon.

Table 1 summarizes the results found in \cite{ref3} and the values of
$\lambda^*$ for the Kuramoto's model in the same motifs. It is
surprising the existence of a possible correlation between
conservation and fitness for synchronization of network
motifs. Namely, the lower $\lambda^*$, the higher the natural
conservation rate. From Table\ \ref{table1}, one observes that: $i)$
$\lambda^{*}_1<\lambda^{*}_2<\lambda^{*}_5$, for the chain-like
motifs; $ii)$
$\lambda^{*}_5>\lambda^{*}_7>\lambda^{*}_8>\lambda^{*}_9$, for motifs
ordered in increasing degree of complexity, i.e., when one can go from
the lower configuration $(5)$ to next one by adding a new link
arbitrarily; and $iii)$
$\lambda^{*}_5>\lambda^{*}_6>\lambda^{*}_8>\lambda^{*}_9$, when
interconnectedness between motif constituents increases. This may
indicate that motifs displaying an improved fitness to develop
cooperative activities are preserved across evolution with a higher
probability.

The results here obtained for topological motifs may be seen on a more
general basis concerning the architecture of real biological and
social networks. It is known that for most of these networks, the
probability of finding cycles (motifs) in their structure is higher
than that expected from a completely random graph with the same
connectivity distribution. For social and biological networks,
friendship or business relationships and natural selection seem to be
at the origin of such cycles. From this perspective, one may assume
that these networks have been shaped during their evolution by some
kind of optimization mechanism that improves on a local level their
ability to develop cooperation. If a new link is created and
afterwards it reveals as not beneficial, then it is removed at later
times.

In summary, we have studied the synchronization of Kuramoto's phase
coupled oscillators for the cases of scale-free networks and
motifs. The results obtained indicate that the architecture of random
SF networks allows synchronization at a small value of the coupling
strength $\lambda$. Additionally, motifs with high interconnectedness
show the lower synchronization thresholds.  Moreover, the results
presented here for the Kuramoto's model hold for other non-linear
dynamical systems, including chaotic oscillators \cite{ref15}. Our
results then lead to a twofold conclusion. On one hand, they suggest
that non-linear mechanisms may be key ingredients for the
understanding of the evolution of networks at a local scale. On the
other hand, as recently unraveled for protein and other biological
networks, the real topology of the systems under analysis is worth
taking into account \cite{alexv}. Finally, we point out that the kind
of models studied here together with the spatial complexity of the
underlying net is relevant for cell biology. A recent example can be
found in the discovery that an ultradian clock (oscillator) shapes
genome expression in the yeast {\em Saccharomyces cerevisiae}
\cite{clock}.

\begin{table}[t]
\begin{tabular}{c}
\epsfig{file=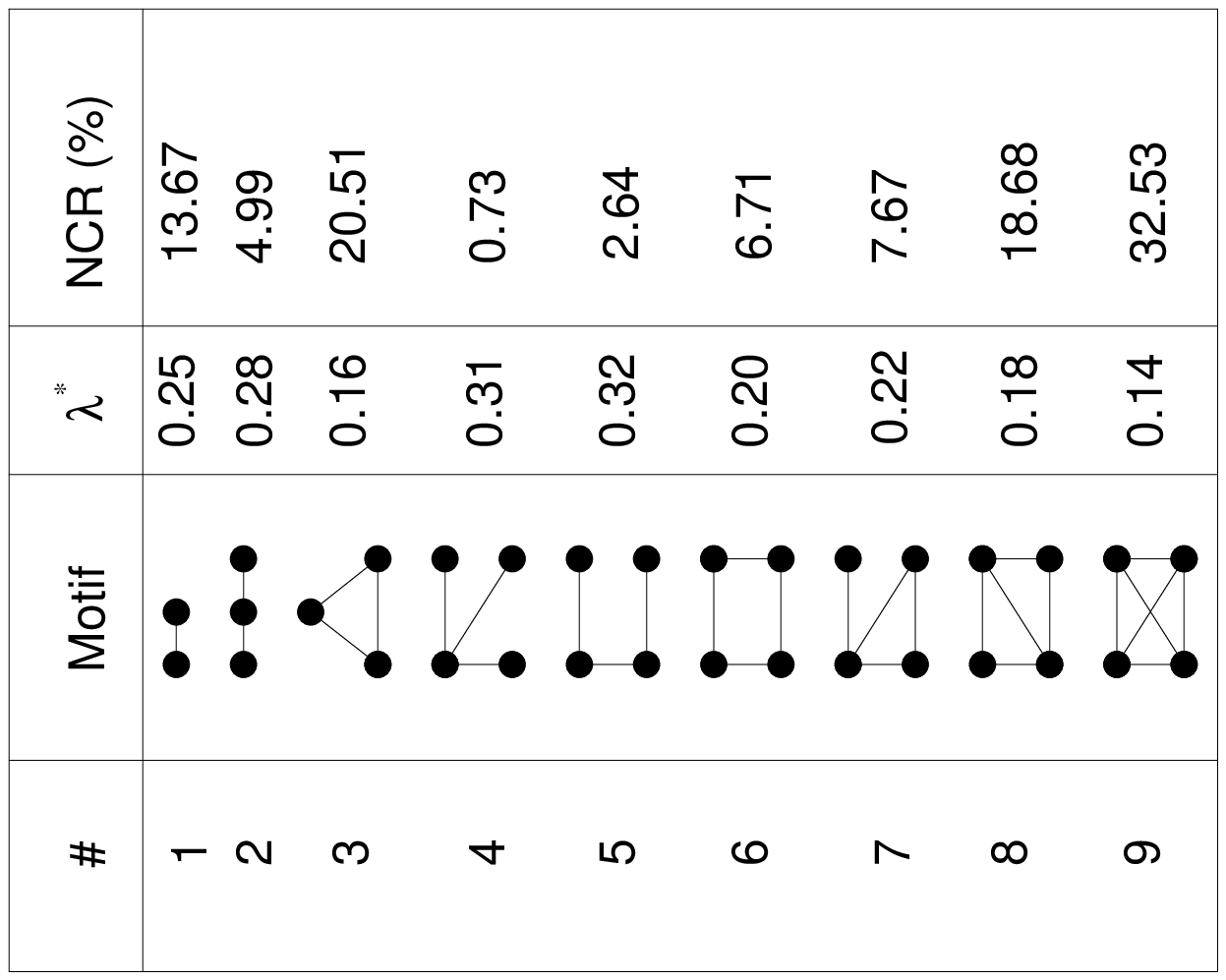,height=4.5in,width=2.8in,angle=-90,clip=1} 
\end{tabular}
\caption{The third column gives the value of the coupling strength
beyond which the motif synchronizes with probability $P_{sync}\ge
\frac{1}{2}$. Natural conservation rates (NCR) reported in the table
are taken from \cite{ref3}. The values reported are sorted in such a
way that for a fixed $N$, the NCR values are in increasing
order. Although not shown due to the lack of NCR values, our
conclusions hold for five-node and higher motifs. In particular, it is
straightforward to conclude from the mean-field description of the
Kuramoto model (all-to-all architecture) that as N increases, the
synchronization threshold decreases.}
\label{table1}
\end{table}

\begin{acknowledgments}
Y.\ M.\ acknowledges financial support from a BIFI research
grant. This work has been partially supported by the Spanish DGICYT
project BFM2002-01798.
\end{acknowledgments}


\begin{thebibliography}{99}

\bibitem{strogatz} S. H. Strogatz, Nature (London) {\bf 410}, 268
(2001).

\bibitem{ref5} H. Jeong, S. P. Mason, A.-L. Barab\'asi, and
  Z. N. Oltvai, Nature (London) {\bf 411}, 41 (2001).

\bibitem{ref6} R. V. Sol\'e, and J. M. Montoya, Proc. R. Soc. London B
  {\bf 268}, 2039 (2001).

\bibitem{pnas} M. E. J. Newman, Proc. Natl. Acad. Sci. U.S.A. {\bf
  98}, 404 (2001).

\bibitem{alexei} A. V\'azquez, R. Pastor-Satorras, and A. Vespignani,
Phys. Rev. E {\bf 65}, 066130 (2002).

\bibitem{ws98} D. J. Watts and H. S. Strogatz, Nature {\bf 393}, 440
  (1998).

\bibitem{newman00} D. S. Callaway, M. E. J. Newman, S. H. Strogatz,
and D. J. Watts, Phys. Rev. Lett. {\bf 85}, 5468 (2000).

\bibitem{pv01a} R. Pastor-Satorras, and A. Vespignani,
Phys. Rev. Lett. {\bf 86}, 3200 (2001).

\bibitem{moreno02} Y. Moreno, R. Pastor-Satorras, and A. Vespignani,
	Eur. Phys. J. B {\bf 26}, 521 (2002).

\bibitem{av03} A. V\'{a}zquez, and Y. Moreno, Phys. Rev. E {\bf 67},
	015101(R) (2003).

\bibitem{moreno03} Y. Moreno, J. B. G\'omez, and A. F. Pacheco,
Phys. Rev. E {\bf 68}, 035103(R) (2003).

\bibitem{sbook} S. H. Strogatz, {\it Nonlinear dynamics and chaos}
(Reading, MA: Perseus Books, Cambridge MA, 1994).

\bibitem{pik} A. Pikovsky, M. Rosenblum, and J. Kurths, {\it
Synchronization: A Universal Concept in Nonlinear Science} (Cambridge
Univ. Press, Cambridge, 2001).

\bibitem{ref3} S. Wuchty, Z. N. Oltvai, and A.-L. Barab\'{a}si, {\it
Nat. Genet.} 35, 176-179 (2003).

\bibitem{k1} Y. Kuramoto, {\it Chemical Oscillations, Waves, and
Turbulence} (Springer, Berlin, 1984).

\bibitem{k2} S. H. Strogatz, Physica D {\bf 143}, 1 (2000).

\bibitem{bar99} A.-L. Barab\'{a}si, and R. Albert, Science {\bf 286},
509 (1999); A.-L. Barab\'{a}si, R. Albert, and H. Jeong, Physica A
{\bf 272}, 173 (1999).

\bibitem{us} Y. Moreno, and A. F. Pacheco, unpublished, preprint
  cond-mat/0401266 (2004).

\bibitem{milo1} S. Itzkovitz, R. Milo, N. Kashtan, G. Ziv and U. Alon,
Phys.Rev. E {\bf 68}, 026127. (2003).

\bibitem{milo2} R. Milo, S. Shen-Orr, S. Itzkovitz, N. Kashtan,
D. Chklovskii and U. Alon, Science {\bf 298}, 824 (2002).

\bibitem{cap} G. Bianconi and A. Capocci, Phys. Rev. Lett. {\bf 90},
078701 (2003).

\bibitem{ref1}J.  Hasty, D. McMillen, and J. J. Collins, {\it Nature}
420, 224-230 (2002).

\bibitem{ref2} H. Kitano, {\it Science} 295, 1662 (2002).

\bibitem{ref15} Barahona, M. and Pecora, L.M. {\it Phys. Rev. Lett.}
89, 054101 (2002).

\bibitem{alexv} A. Vespignani, {\it Nat. Genet.} 35, 118 (2003).

\bibitem{clock} R. R. Klevecz, J. Bolen, G. Forrest, and D. B. Murray,
  {\it Proc. Natl. Acad. Sci. USA} {\bf 101}, 1200 (2004).

\end{thebibliography}
\end{document}